 \documentclass[twocolumn,prl,aps,psfig,preprintnumbers,superscriptaddress]{revtex4}
\usepackage{graphicx}
\usepackage{epstopdf}
\usepackage{float}
\usepackage{color}
\usepackage{amsmath}

\begin{document}

\title{Magnetic detwinning and biquadratic magnetic interaction in EuFe$_2$As$_2$ revealed by $^{153}$Eu NMR}
\author{Q.-P. Ding}
\affiliation{Ames Laboratory, U.S. DOE, and Department of Physics and Astronomy, Iowa State University, Ames, Iowa 50011, USA}
\author{N. S. Sangeetha}
\affiliation{Ames Laboratory, U.S. DOE, and Department of Physics and Astronomy, Iowa State University, Ames, Iowa 50011, USA}
\author{W. R. Meier}
\affiliation{Ames Laboratory, U.S. DOE, and Department of Physics and Astronomy, Iowa State University, Ames, Iowa 50011, USA}
\author{M. Xu}
\affiliation{Ames Laboratory, U.S. DOE, and Department of Physics and Astronomy, Iowa State University, Ames, Iowa 50011, USA}
\author{S. L. Bud'ko}
\affiliation{Ames Laboratory, U.S. DOE, and Department of Physics and Astronomy, Iowa State University, Ames, Iowa 50011, USA}
\author{P. C. Canfield}
\affiliation{Ames Laboratory, U.S. DOE, and Department of Physics and Astronomy, Iowa State University, Ames, Iowa 50011, USA}
\author{D. C. Johnston}
\affiliation{Ames Laboratory, U.S. DOE, and Department of Physics and Astronomy, Iowa State University, Ames, Iowa 50011, USA}
\author{Y. Furukawa}
\affiliation{Ames Laboratory, U.S. DOE, and Department of Physics and Astronomy, Iowa State University, Ames, Iowa 50011, USA}

\date{\today}

\begin{abstract} 
     In the nematic state of iron-based superconductors, twin formation often obscures the intrinsic, anisotropic, in-plane physical properties.
    Relatively high in-plane external magnetic fields $H_{\rm ext}$ greater than the typical lab-scale magnetic fields 10--15 T are usually required to completely detwin a sample.   
     However, recently a very small in-plane $H_{\rm ext} \sim$ 0.1 T was found to be sufficient for detwinning the nematic domains in EuFe$_2$As$_2$. 
    To explain this behavior, a microscopic theory based on biquadratic magnetic interactions between the Eu and Fe spins has been proposed.
      Here, using $^{153}$Eu nuclear magnetic resonance (NMR) measurements below the Eu$^{2+}$ ordering temperature, we show experimental evidence of the detwinning under small in-plane $H_{\rm ext}$.
      Our NMR study also reveals the evolution of the angles between the Eu and Fe spins during the detwinning process, which provides the first experimental evidence for the existence of biquadratic coupling in the system.
\end{abstract}

\maketitle

    The discovery of iron-based superconductors (IBSCs) triggered intense activity in the research field of so-called \lq\lq{}electronic nematicity\rq\rq{} which can be identified in the orthorhombic structure where the magnitude of the electronic anisotropy cannot be simply explained by the effect of the orthorhombic lattice distortion \cite{Johnston2010,Canfield2010,Stewart2011}. 
   Measurements of the the intrinsic in-plane physical properties to clarify the characteristics of  electronic nematicity are usually hampered by twin formation in nematic state.  
   Up to now, two distinct methods have been mainly employed to detwin the crystals: The application of uniaxial strain \cite{Fisher2011,Tanatar2010,Chu2010,Ying2011,Dhital2012,Yi2011,Kim2011,Kissikov2017,Kissikov2018} and the application of an in-plane external magnetic field ($H_{\rm ext}$) \cite{Chu20102,Ruff2012}.
   However, both methods may obscure the intrinsic properties of the compounds.
   For example, uniaxial strain may change the nematic and magnetic transition temperatures \cite{Dhital2012}.
   In addition, relatively high $H_{\rm ext}$  is required to complete the detwinning (e.g., $\sim$27 T for Ba(Fe$_{1-x}$Co$_x$)$_2$As$_2$ \cite{Ruff2012}),
although a change in the relative twin population can be produced by a smaller in-plane $H_{\rm ext}$ (e.g., $\sim$14 T for Ba(Fe$_{1-x}$Co$_x$)$_2$As$_2$  \cite{Chu20102}).
    
   Recently, EuFe$_2$As$_2$ has attracted much attention since a small in-plane $H_{\rm ext}$, less than 0.5 T, is enough to complete the detwinning.
    Different from most non-rare-earth bearing so-called 122 IBSCs, EuFe$_2$As$_2$ exhibits two magnetic phase transitions \cite{Jeevan2008,Tereshima2009}.
    The first magnetic ordering  state below $\sim$ 189 K is the stripe-type antiferromagnetic (AFM) state due to the Fe moments with a concomitant first-order structural phase transition to a low-temperature  orthorhombic structure corresponding to the nematic transition.
The second one below 19 K is associated with  Eu$^{2+}$ moments, making an A-type AFM structure where the Eu moments are ferromagnetically aligned in the $ab$ plane but the moments in adjacent layers along the $c$ axis are antiferromagnetically aligned \cite{Jeevan2008}. 
    A realignment of the twinning structure by in-plane $H_{\rm ext}$ has been first observed in a single-crystal neutron diffraction (ND) measurement \cite{Xiao2010}.
    A detailed study of EuFe$_2$As$_2$ using resistivity, thermal-expansion, magnetostriction, magnetoresistance, magneto-optical, and magnetization measurements shows that a very low in-plane $H_{\rm ext}$ of only $\sim$ 0.1 T is sufficient for detwinning below the Eu$^{2+}$ ordering temperature \cite{Zapf2014}. 
    Furthermore, the detwinning effects  remain up to the nematic transition temperature well above  the Eu$^{2+}$ ordering temperature even after $H_{\rm ext}$ is switched off \cite{Zapf2014}. 
    Thus, this detwinning provides a unique way to study the low-temperature in-plane physical properties of EuFe$_2$As$_2$.

    A recent microscopic theory with a biquadratic coupling between the Eu and Fe spins has been proposed to explain the detwinning process in this compound \cite{Maiwald2018}.
    According to the theory,  the detwinning can be initiated by the application of a small in-plane $H_{\rm ext}$ less than 0.1 T and then a complete detwinning can be attained at the first detwinning magnetic field $H_1$ around 0.3--0.5 T where only one domain remains. 
  With further application of $H_{\rm ext}$, a part of the domain spontaneously rotates by 90$^{\circ}$ where the angle between $H_{\rm ext}$ and the Eu spins is proposed to change from 55$^\circ$ to 25$^\circ$ due to the existence of the biquadratic magnetic interaction \cite{Maiwald2018} whereas a simple spin-flip is expected without the interaction \cite{Zapf2014}.
   Furthermore, the theory proposes that the population of the new domain increases with increasing $H_{\rm ext}$ and the new domain dominates at the second detwinning magnetic field $H_2$ around 1 T.   
   Therefore, it is important to investigate the details of how the angles between Eu and Fe spins change during the detwinning processes so as to test the theory where the biquadratic magnetic interaction has been proposed to play an important role.
 
   To elucidate the evolution of the detwinning process in EuFe$_2$As$_2$,  we used nuclear magnetic resonance (NMR) which is one of the suitable experimental techniques to provide  the required information from a microscopic point of view. 
  Here we have succeeded in observing $^{153}$Eu NMR signals in the magnetically-ordered state of EuFe$_2$As$_2$ and also in determining the details of the change of the angles between Eu and Fe spins with the application of $H_{\rm ext}$. 
   Our NMR results provide direct evidence of the magnetic detwinning for in-plane $H_{\rm ext}$ and also, for the first time,  of the existence of a biquadratic magnetic interaction as proposed by the theory \cite{Maiwald2018}.




\begin{figure}[tb]
\includegraphics[width=\columnwidth]{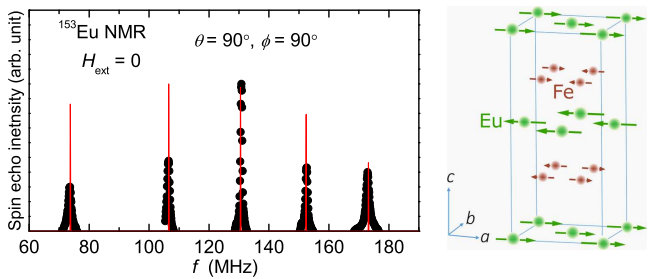} 
\caption{ $^{153}$Eu-NMR spectrum for $H_{\rm ext}$ = 0 at 4.3 K. The red lines are the calculated $^{153}$Eu-NMR spectrum. The right figure shows the spin structure of EuFe$_2$As$_2$ in the orthorhombic phase under zero $H_{\rm ext}$.
  }
\label{fig:NMR}
\end{figure}   

   Details of the sample preparation and NMR measurements  are described in supplementary material \cite{experiment}. 
   Figure  \ \ref{fig:NMR} shows the $^{153}$Eu NMR spectrum in zero $H_{\rm ext}$ at 4.3 K in the magnetically-ordered state. 
  The observation of the $^{153}$Eu zero-field NMR signals clearly evidences that the magnetic moments of Eu 4$f$ electrons order in the magnetic state. 
   The peak positions of the spectrum can be explained by the combination of a large Zeeman interaction (${\cal H}_{\rm Z}$) due to $H_{\rm ext}$ [for the present case, an internal magnetic induction ($B_{\rm int}$) at the Eu site] and a small quadrupole interaction (${\cal H}_{\rm Q}$),  whose nuclear spin Hamiltonians are given by   
${\cal H}_{\rm Z} = -\gamma_{\rm n}\hbar  {\bf I} \cdot {\bf B_{\rm int}} $, and $ {\cal H}_{\rm Q} = \frac{h\nu_{\rm Q}}{6}[3I_Z^2-I^2 + \frac{1}{2}\eta(I_+^2 +I_-^2)]$, respectively. 
Here $h$ is Planck's constant, $\nu_{\rm Q}$ is the nuclear quadrupole frequency defined by $\nu_{\rm Q} = 3e^2QV_{ZZ}/20h$  where $Q$ is the electric quadrupole moment of the Eu nucleus, $V_{ZZ}$ is the electric field gradient (EFG) at the Eu site  in the coordinates of the principal $X$, $Y$, and $Z$ axes of the EFG, and $\eta$ is the asymmetry parameter of the EFG defined by $\frac{V_{XX} -V_{YY}}{V_{ZZ}}$ with $|V_{ZZ}|$$\geq$$|V_{YY}|$$\geq$$|V_{XX}|$  \cite{Slichterbook}.
 In this case, the resonance frequency $f$ for the transition from $I_z$ = $m$ to $m-1$ is given within first-order perturbation theory by \cite{Abragambook}
 \begin{eqnarray*}
 \lefteqn{ f(m \leftrightarrow m-1)}  \\
& =&   \nu_0 + \frac{1}{2}\nu_{\rm Q}(m-\frac{1}{2})(3\cos^2\theta-1+\eta \sin^2\theta \cos2\phi ). 
  \label{eq:3}
  \end{eqnarray*} 
     Here $\frac{\gamma_{\rm N}}{2 \pi}B_{\rm int}  = \nu_0$, and    $\theta$ and $\phi$ are the polar and azimuthal angles between  the $Z$ axis of the EFG and the direction of $B_{\rm int}$, respectively, where the quantization axis ($z$ axis) for the Zeeman interaction points along the $B_{\rm int}$ direction. 
    Thus, from spectrum measurements, especially from the spacings between the lines, one can estimate the angles $\theta$ and $\phi$ which provide important information about the direction of $B_{\rm int}$ with respect to the EFG coordinate system. 
   The observed line positions were well reproduced by the calculation, as shown by the red lines in Fig. \ \ref{fig:NMR}(a), with the parameters $|B_{\rm int}^{\rm Eu}|$  = 27.0 T, $\nu_{\rm Q}$ = 40.2  MHz, $\eta$ = 0.25,  $\theta = 90^\circ$, and $\phi = 90^\circ$. 
    The sign of $B_{\rm int}$ at the Eu site is considered to be negative because $B_{\rm int}$ mainly originates from core polarization from 4$f$ electrons and is oriented in a direction opposite to that of the Eu ordered moments \cite{Freeman1965}.
    Comparable values of $B_{\rm int}$ at Eu sites were reported in similar compounds such as the helical antiferromagnets EuCo$_2$As$_2$ ($B_{\rm int}$ = $-$25.75 T) \cite{Ding2017} and EuCo$_2$P$_2$  ($B_{\rm int}$ = $-$27.5 T) \cite{Higa2017} and the A-type antiferromagnet EuGa$_4$ ($B_{\rm int}$ = $-$27.08 T) \cite{Yogi2013}.
     Since the Eu ordered moments in those compounds are close to 7 $\mu_{\rm B}$ as well as in EuFe$_2$As$_2$, the similar value of $B_{\rm int}$ = $-$27.0 T  suggests that  the hyperfine field induced by Fe moments at the Eu site is negligible and the dominant contribution to $B_{\rm int}$ at the Eu site is from the Eu ordered moments.
      To obtain more information about the principal axes of the EFG at the Eu site, we have calculated the EFG using a point-charge model.  
      We found that the $X$, $Y$, and $Z$ axes correspond to the $b$, $a$, and $c$ axes in the orthorhombic structure, respectively, and $\eta =0.28$ whose value is close to $\eta =0.25$ estimated from fitting the spectrum.
    Therefore, the results of both $\theta = 90^\circ$ and $\phi = 90^\circ$ indicate that $B_{\rm int}$ is parallel or antiparallel to the $a$ axis, which is consistent with the magnetic structure under zero  $H_{\rm ext}$  shown in Fig. \ref{fig:NMR} determined by the ND measurements \cite{Jeevan2008}.

\begin{figure}[tb]
\includegraphics[width=\columnwidth]{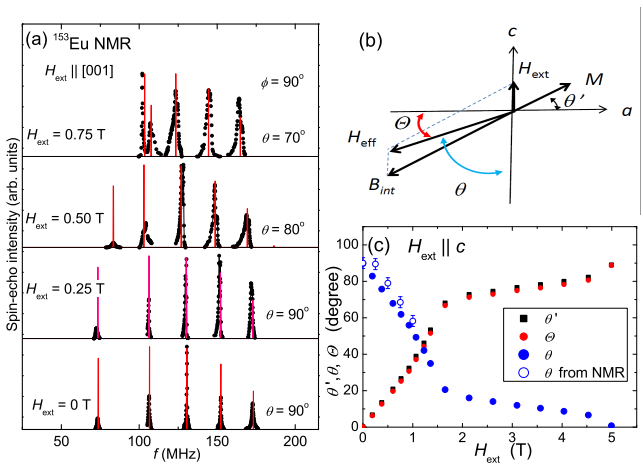} 
\caption{(a) $H_{\rm ext}$ dependence of the $^{153}$Eu-NMR spectra of EuFe$_2$As$_2$ for $H_{\rm ext}$$\parallel$$c$ at $T$ = 4.3 K.
The red lines are calculated spectra with different values of $\theta$ and $H_{\rm eff}$ under different $H_{\rm ext}$ without changing other parameters: $B_{\rm int}^{\rm Eu}$  = $-$27.0 T, $\nu_{\rm Q}$ = 40.2  MHz, $\eta$ = 0.25,  and $\phi = 90^\circ$.    
     (b) Schematic view of the configuration for $\theta$ and the canting angles of  $\theta^{\prime}$ and $\Theta$  between the magnetization and the quantization axis of Eu nucleus, respectively,  from the $ab$ plane in $H_{\rm ext}$$\parallel$$c$. 
(c) $H_{\rm ext}$$\parallel$$c$ dependence of the angles $\theta$, $\theta^{\prime}$ and $\Theta$ estimated from the magnetization data at $T$ = 5 K \cite{Jiang2009} and $\theta$ estimated from $^{153}$Eu NMR spectrum measurements.  
}
\label{fig:NMR_Hc}
\end{figure}

      Figure \ref{fig:NMR_Hc}(a)  shows the dependence of the $^{153}$Eu spectra on $H_{\rm ext}$ applied parallel to the $c$ axis ($H_{\rm ext}$$\parallel$$c$)   in the AFM state at 4.3 K.
      In the AFM state, one expects a splitting of the NMR line when $H_{\rm ext}$ is applied along the magnetic easy axis, while only a shift of the NMR line without splitting is expected when $H_{\rm ext}$ is applied perpendicular to the magnetic easy axis, for $H_{\rm ext}$ smaller than the magnetocrystalline anisotropy field. 
     Since the magnetic easy axis is parallel to the $a$ axis, we do not expect the splitting of the line for $H_{\rm ext}$$\parallel$$c$, as actually observed.  
       The effective field ($H_{\rm eff}$) at the Eu site  is the vector sum of $\bf{B}_{\rm int}$ and $\bf{H_{\rm ext}}$,  i.e., $|$$\bf{H}_{\rm eff}$$|$ = $|$$\bf{B}_{\rm int}$ + $\bf{H_{\rm ext}}$$|$. 
     Therefore, utilizing  the canting angle $\theta^{\prime}$ of the Eu ordered moment from the $a$ axis to the $c$ axis [see Fig. \ref{fig:NMR_Hc}(b)], the $H_{\rm eff}$ can be written as $H_{\rm eff}$  =   $\sqrt{\mathstrut H_{\rm ext}^2 + B_{\rm int}^2 + 2H_{\rm ext}B_{\rm int}{\rm sin}\theta^{\prime}}$. 
  Here $\theta^{\prime}$ can be calculated from magnetization data  since $\theta^{\prime}$ = sin$^{-1}$($M/M_{\rm s})$ where $M_{\rm s}$ is the saturation of magnetization.
   Figure \ref{fig:NMR_Hc}(c) shows the calculated $H_{\rm ext}$ dependence of $\theta^{\prime}$  for $H_{\rm ext}$$\parallel$$c$  from the magnetization data reported in Ref. \cite{Jiang2009}  where we used $M_{\rm s}$ = 7 $\mu_{\rm B}$.
    It is noted that the quantization axis for the Zeeman interaction points in the direction of $H_{\rm eff}$ which is in general not the same as that of $B_{\rm int}$ as shown in Fig. \ref{fig:NMR_Hc}(b). 
    Using $B_{\rm int}$ = $-$27 T, we calculated the $H_{\rm ext}$ dependence of $\Theta$ and $\theta$ and found that the difference  between $\Theta$ and $\theta^{\prime}$ is less than 2 degrees due to the large value of $B_{\rm int}$ with respect to $H_{\rm ext}$ for our experimental region, as shown in Fig.  \ref{fig:NMR_Hc}(c).
   Therefore, we approximate  $\theta^{\prime}$ $\sim$ $\Theta$ and used this approximation to estimate $\theta$ shown in Fig.  \ref{fig:NMR_Hc}(c) which can be compared with the results of NMR measurements.  

   As shown by the vertical red lines in Fig.  \ref{fig:NMR_Hc}(a), the observed spectra for $H_{\rm ext}$$\parallel$$c$ are well reproduced by changing $\theta$ and ${H}_{\rm eff}$ with other parameters unchanged. 
   The $H_{\rm ext}$ dependence of $\theta$ determined from the NMR spectra is in good agreement with $\theta$ estimated from the magnetization data [see Fig.  \ref{fig:NMR_Hc}(c)].
  Thus we conclude that the Eu ordered moments change the direction from the $a$ axis to the $c$ axis continuously with increasing $H_{\rm ext}||c$, and eventually will point to the $c$ axis at higher $H_{\rm ext}$.
   Finally it is noted that we observed small splitting with broadening of each line under $H_{\rm ext}||c$. 
   This could be due to a slight misalignment of the crystal or a small magnetic induction at the Eu site produced by the Fe ordered moments. 
 
 \begin{figure*}[tb]
\includegraphics[width=\textwidth]{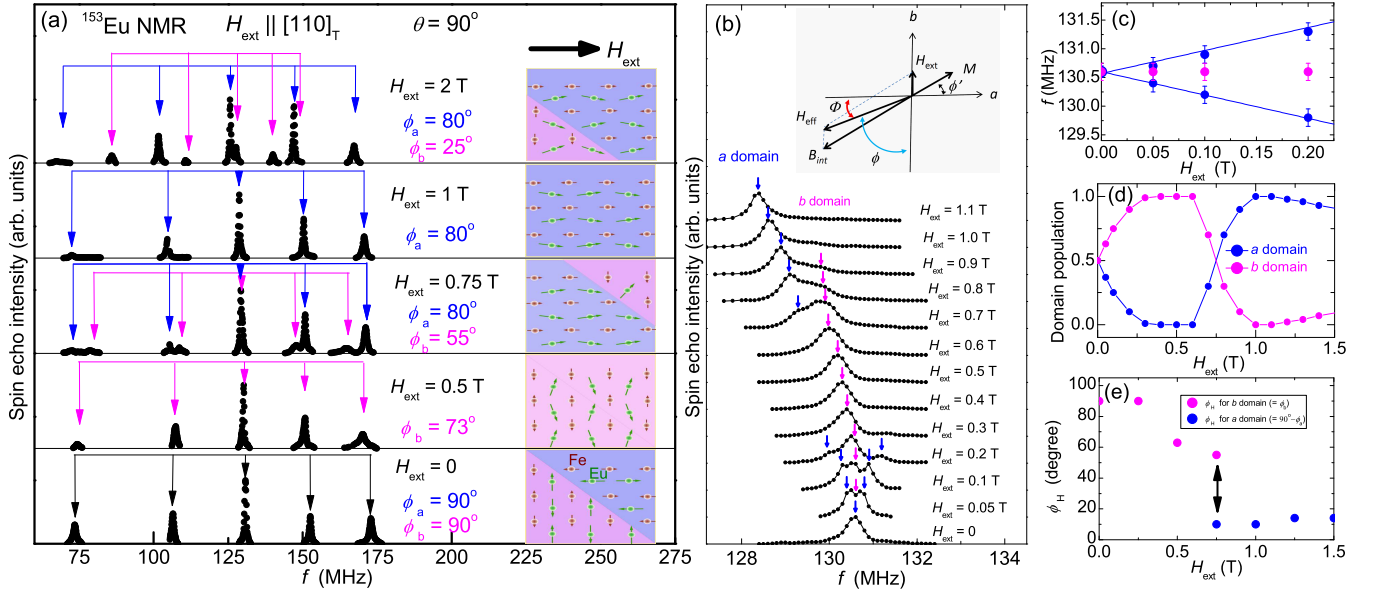} 
\caption{(a) $H_{\rm ext}$ dependence of the $^{153}$Eu-NMR spectrum for $H_{\rm ext}$$\parallel$[110]$_{\rm T}$ at $T$ = 4.3 K.
The calculated positions of the spectra for the $b$ and $a$ domains are shown by the arrows in magenta and blue, respectively.
The insets show a sketch of the spin directions of Eu (green) and Fe (brown) for the $a$ (blue) and $b$ (magenta) domains.
(b) $H_{\rm ext}$  dependence of the central peak of the $^{153}$Eu-NMR spectrum  for $H_{\rm ext}$$\parallel$[110]$_{\rm T}$.  
 The inset shows a schematic view of the configuration for the angles $\phi$, $\phi^{\prime}$ and $\Phi$ in the $ab$ plane for the case of $H_{\rm ext}$$\parallel$$b$. 
(c) $H_{\rm ext}$ dependence of $f$ of the central transition line for an $a$ domain (blue) and a $b$ domain (magenta).  The solid lines are linear fits by $f$ = $\nu_0$ $\pm$ $\alpha$$H_{\rm eff}$ with $\alpha$ = 4 MHz/T.    
(d) $H_{\rm ext}$ dependence of domain population.
(e) $H_{\rm ext}$ dependence of the angle $\phi_{\rm H}$ between $H_{\rm ext}$ and the Eu spins for the $a$ and $b$ domains. 
}
\label{fig:ab}
\end{figure*}    

   When $H_{\rm ext}$ is applied in the $ab$ plane, the $^{153}$Eu NMR spectra show quite different behavior as shown in  Figs. \ref{fig:ab}(a) and 3(b), which evidence control of the domain populations with $H_{\rm ext}$ as will be discussed below in detail. 
 Here we applied $H_{\rm ext}$ parallel to [110] in the tetragonal structure ([110]$_{\rm T}$). The notation of the tetragonal structure is rotated by 45$^{\circ}$ along the $c$ axis with respect to the low-temperature orthorhombic structure.
   Thus the [110]$_{\rm T}$ direction is parallel to the $a$ (the direction of Fe spins) or $b$ (perpendicular to the direction of Fe spins) axes in the orthorhombic structure [see Fig. \ \ref{fig:NMR}(b)].

    With the application of small $H_{\rm ext}$, we found that  each line splits into three lines as typically shown in Fig. \ref{fig:ab}(b).
    This clearly evidences the existence of two domains. 
    For $H_{\rm ext}$$\parallel$$a$, one expects the symmetric splitting of the line because $B_{\rm int}$ is parallel or antiparallel to $H_{\rm ext}$, that is, $H_{\rm eff}$ = $B_{\rm int}$ $\pm$ $H_{\rm ext}$. 
    The $H_{\rm ext}$ dependence of the splitted peak frequencies is shown in Fig. \ref{fig:ab}(c) where the absolute values of the slopes for the $H_{\rm ext}$  dependence are $\sim$4 MHz/T, which is close to the $\frac{\gamma_{\rm N}}{2\pi}$ value of the Eu nucleus. 
     On the other hand, when $H_{\rm ext}$ is parallel to the $b$ axis, any shift in the position would be very small.
    This is actually observed as shown in Fig. \ref{fig:ab}(c), thus the line can be assigned to be from the domain with $H_{\rm ext}$$\parallel$$b$ axis.
     Hereafter a domain with the $a$ or $b$ axis along $H_{\rm ext}$ is defined as an ``$a$ domain" or ``$b$ domain", respectively.  

    With increasing $H_{\rm ext}$, the signal intensity of the splitted peaks is reduced, indicating that the population of the $a$ domain decreases.  
    Around 0.3--0.6 T, we observed a set of lines only from the $b$ domain, as typically shown at the second  panel from the bottom in Fig. \ref{fig:ab}(a). 
      This magnetic field of 0.3 T corresponds to the first detwinning field $H_1$ whose value is consistent with the results from a recent magnetization measurement \cite{Maiwald2018} and is close to $\sim$0.4 T estimated from the ND measurement \cite{Xiao2010}, although it is  slightly higher than $\sim$0.1 T reported  from the magneto-optical measurements \cite{Zapf2014}.
     This detwinning process has been explained as follows: since the $b$ domain (i.e., $H_{\rm ext}$$\perp$Eu ordered moments) is lower in energy than the $a$ domain ($H_{\rm ext}$ parallel or antiparallel to the Eu ordered moments), the $b$ domain can push the $a$ domain out once the energy difference between the domains overcomes the domain boundary pinning energy \cite{Zapf2014}. 

    It is noted that from the spacings between NMR-spectrum lines in Fig. \ref{fig:ab}(a), one can estimate the angle $\phi$ for the $b$ domain ($\phi_{\rm b}$) which corresponds to the angle between $H_{\rm eff}$ and the $b$ axis. 
  Here $\phi$ can be approximated by 90$^{\circ}$-$\phi^{\prime}$ since the angle $\Phi$ is close to the Eu spin\rq{s} canting angle ($\phi^{\prime}$) in the $ab$ plane  [see the inset of Fig. \ref{fig:ab}(b)],  similar to the case of $H_{\rm ext}$$\parallel$$c$.
The $\phi$ value in the $b$ domain decreases from 90$^\circ$ at $H_{\rm ext}$ = 0 to 73$^\circ$ at $H_{\rm ext}$ = 0.5 T, corresponding to the Eu spin canting in the $ab$ plane, as illustrated in Fig. \ \ref{fig:ab}(a).
   This phase with dominant $b$ domain is found to have an $H_{\rm ext}$ range of 0.3$-$0.6 T, which is consistent with the results reported by other techniques \cite{Xiao2010,Zapf2014,Maiwald2018}.


    When $H_{\rm ext}$ is increased to $\sim$0.7 T, another set of lines appears as shown in the middle panel of Fig. \ \ref{fig:ab}(a). 
   At the similar $H_{\rm ext}$ of 0.7 T,  the magnetization was reported to show a sudden increase \cite{Jiang2009,Zapf2014}, which was either ascribed to a  metamagnetic transition 
\cite{Jiang2009} or a spin-flip transition \cite{Zapf2014}. 
   However, the recent magnetostriction and magnetotransport measurements suggested that the jump in magnetization is associated with the reorientation of domains \cite{Maiwald2018}.
        According to Maiwald $et~ al$.,  if the biquadratic coupling is finite, the system changes domains from $b$-type to $a$-type where the angle between the Eu spins and $H_{\rm ext}$ ($\phi_{\rm H}$) changes discontinuously from  $\phi_H > \pi/4$ in the $b$ domain to  $\phi_H < \pi/4$ in the $a$ domain \cite{Maiwald2018}. 
     This is due to the biquadratic magnetic interaction which makes the angle ($\beta$)  between the Eu and Fe spins small because the energy is proportional to the square of cosine of $\beta$ ($E$ $\propto$ $-$cos$^2\beta$) \cite{Maiwald2018}. 
    Since the Fe spins point along the $a$ axis, the retwinning with discontinuous change in $\phi_{\rm H}$ due to the domain rotation from the $b$ domain to the $a$ domain can be possible, when the energy gain exceeds the pinning energy of the domain boundary \cite{Maiwald2018}.

      Our NMR spectrum at $H_{\rm ext}$ = 0.75 T shows the superposition of two spectra originating from the $b$ domain with  $\phi_b$ = 55$^\circ$  and the $a$ domain with  $\phi_a$ = 80$^\circ$, clearly evidencing the  reappearance of the $a$ domain, as depicted in right-hand sides of the panels in Fig. \ \ref{fig:ab}(a).
     From the values of $\phi_b$ (=$\phi_H$) and $\phi_a$ (=90$^\circ$--$\phi_H$), it is found that  the $\phi_H$ changes discontinuously from $\sim$55$^{\circ}$ in the $b$ domain to $\sim$10$^{\circ}$ in the $a$ domain at 0.75 T [see Fig.\ \ref{fig:ab}(e)]. 
   These values of $\phi_{\rm H}$ are in good agreement with  55$^{\circ}$ and  25$^{\circ}$ estimated from the theoretical calculation \cite{Maiwald2018}.
   Thus we conclude that our NMR data provide direct evidence for the existence of the biquadratic coupling in EuFe$_2$As$_2$. 

 
   The spectral weight moves from the $b$ domain to the $a$ domain when $H_{\rm ext}$ is increased from 0.7 T to around 1 T, as shown in Figs. \ \ref{fig:ab}(a) and \ \ref{fig:ab}(b), showing that the population of the $b$ domain decreases while that of the $a$ domain increases.
   The $b$ domain nearly vanishes around 1 T, leaving only the $a$ domain at this $H_{\rm ext}$, as shown in Fig. \ \ref{fig:ab}(a), which corresponds to  the second detwinning field $H_2 \sim$ 1 T. This is also consistent with the theory.  

    Finally, it is interesting to point out that when $H_{\rm ext}$ is further increased, as shown by the NMR spectra for $H_{\rm ext}$ = 2 T in Fig. \ \ref{fig:ab}(a), two sets of NMR spectra originating from both $a$ and $b$ domains appear again. 
    This indicates that the $b$ domain recovers at this field, as illustrated in Fig. \ \ref{fig:ab}(a).
   Such retwinning has been proposed from the theory \cite{Maiwald2018}. 
   However, since the twinning is due to the spin-flop of Fe ordered moments \cite{Maiwald2018}, the retwinning is expected above at least 10 T, which cannot be simply applied to the explanation of our observation.  
     At present, although we do not have a clear idea why the retwinning starts to appear above $\sim$1.2 T, it could be due to a sort of local effect such as Fe-ion defects producing less energy to rotate the domain since the relative population of the $b$ domain is nearly independent from 2 T to  the measured highest $H_{\rm ext}$ of 6 T. 
Further studies under higher $H_{\rm ext}$ are required to clarify this point.

    In summary, we reported the results of $^{153}$Eu-NMR measurements on an EuFe$_2$As$_2$ single crystal, which provides evidence for detwinning under a small in-plane $H_{\rm ext}$ from a microscopic point of view.  
   The two magnetic detwinning states around $H_1$ $\sim$ 0.3 T and  $H_2$ $\sim$ 1 T for $H_{\rm ext}$$\parallel$[110]$_{\rm T}$ were revealed from the NMR spectrum measurements. 
   The evolution of the angles between Eu and Fe spins during the detwinning process was also determined, which provides the first experimental evidence for the existence of the biquadratic coupling in the system.
    From this point of view, it is interesting that  higher superconducting transition temperatures ($T_{\rm c}$) for bulk IBSCs have been observed in rare-earth bearing 1111 \cite{Ren2008} and 122 \cite{Lv2011} systems. 
  Since  our results indicate that the biquadratic coupling between rare-earth and iron moments exists and has a critical role, the enhancement of $T_{\rm c}$ could be related to the biquadratic coupling  in those systems. 
  Further studies in view of the biquadratic coupling are suggested to understand the origin of the high $T_{\rm c}$ in those systems,  which may also  provide some clues about the origin of superconductivity in IBSCs where the biquadratic coupling between the Fe spins is considered to play an important role  \cite{Yaresko2009, Wysocki2011}.

The research was supported by the U.S. Department of Energy (DOE), Office of Basic Energy Sciences, Division of Materials Sciences and Engineering. Ames Laboratory is operated for the U.S. DOE by Iowa State University under Contract No. DE-AC02-07CH11358. W. R. M. was supported by the Gordon and Betty Moore Foundation’s EPiQS Initiative through Grant No. GBMF4411.

\end{document}